\documentclass{article}

\usepackage{arxiv}

\usepackage[utf8]{inputenc} 
\usepackage[T1]{fontenc}    
\usepackage{hyperref}       
\usepackage{url}            
\usepackage{booktabs}       
\usepackage{amsfonts}       
\usepackage{nicefrac}       
\usepackage{microtype}      
\usepackage{lipsum}
\usepackage{graphicx}
\usepackage{array}
\usepackage{gensymb}
\usepackage{tabularx}
\usepackage{float}
\usepackage{amsmath}
\graphicspath{ {./images/} }

\title{Enhancing Diabetic Retinopathy Diagnosis: A Lightweight CNN Architecture for Efficient Exudate Detection in Retinal Fundus Images}

\author{
 Mujadded Al Rabbani Alif \\
  Department of Computer Science\\
  Huddersfield University\\
  Queensgate, Huddersfield HD1 3DH, UK \\
  \texttt{m.alif@hud.ac.uk} \\
}

\begin{document}
\maketitle
\begin{abstract}
Retinal fundus imaging plays an essential role in diagnosing various stages of diabetic retinopathy, where exudates are critical markers of early disease onset. Prompt detection of these exudates is pivotal for enabling optometrists to arrest or significantly decelerate the disease progression. This paper introduces a novel, lightweight convolutional neural network architecture tailored for automated exudate detection, designed to identify these markers efficiently and accurately. To address the challenge of limited training data, we have incorporated domain-specific data augmentations to enhance the model's generalizability. Furthermore, we applied a suite of regularization techniques within our custom architecture to boost diagnostic accuracy while optimizing computational efficiency. Remarkably, this streamlined model contains only 4.73 million parameters—a reduction of nearly 60\% compared to the standard ResNet-18 model, which has 11.69 million parameters. Despite its reduced complexity, our model achieves an impressive F1 score of 90\%, demonstrating its efficacy in the early detection of diabetic retinopathy through fundus imaging.
\end{abstract}


\section{Introduction}
The World Health Organization's 2019 World Vision Report delineates a pressing global health issue: approximately 2.2 billion individuals worldwide are afflicted with some form of visual impairment \cite{who_vision2020}. Critically, the report suggests that up to 1 billion of these cases could have been averted or minimized through early diagnostic measures and timely medical interventions. Diabetic Retinopathy (DR), as identified in the literature, is the predominant cause of vision loss among adults of working age in developed nations \cite{abramoff2010retinal, ciulla2003diabetic}. Retinal imaging has become an indispensable diagnostic modality in contemporary ophthalmology \cite{keane2014retinal}. This non-invasive method provides a comprehensive view of the retina by capturing detailed fundus photographs. Such imaging is instrumental in the thorough assessment of retinal microvascular health \cite{liew2008retinal}, revealing critical insights that surpass those of conventional clinical examinations \cite{shen2020modeling}. The analytical evaluation of these images enables healthcare professionals to diagnose well-established ocular diseases and identify the early markers of progressive conditions like DR. The capacity for early detection through retinal imaging is vital, facilitating prompt medical interventions that can substantially slow the progression of the disease \cite{sisson2014analysis}. This capability is crucial for advancing preventative ophthalmic care and enhancing patient outcomes.

The advent of Deep Learning (DL) technologies, especially Convolutional Neural Networks (CNNs), has catalyzed transformative advances in the field of computer vision \cite{wang2023beyond, albawi2017understanding, voulodimos2018deep, hussain2023and}. CNNs, a dynamic subset of DL within artificial intelligence, are particularly adept at autonomously extracting pertinent features from image data, thereby circumventing the extensive manual feature engineering required by traditional machine learning approaches. However, the deployment of CNNs involves substantial computational demands to achieve optimal functionality \cite{hussain2023child}. In the context of Diabetic Retinopathy (DR), the appearance of exudates during the non-proliferative stage presents a critical diagnostic marker. These lesions, resulting from the compromised integrity of retinal pericytes, lead to enhanced vascular permeability and the subsequent leakage of plasma constituents like lipoproteins and proteins. These components accumulate within the retinal layers, appearing as distinct yellowish deposits on fundus images \cite{cusick2003histopathology, chowdhury2002role}. The development and refinement of automated algorithms for the detection of exudates are crucial, offering substantial potential to elevate early DR diagnosis in clinical settings. Such technological advancements could enable clinicians to execute timely interventions, thus curtailing the progression of the disease and aiding in the preservation of vision.

This study aims to develop and assess a lightweight convolutional neural network architecture tailored for the detection of retinal exudates, focusing on its effectiveness compared to the conventional ResNet-18 model. Given the increasing importance of efficient and accurate diagnostic tools in ophthalmology, it is essential to evaluate how this novel model performs in terms of accuracy, computational efficiency, and generalizability. This analysis will help determine the model's capability in identifying early signs of diabetic retinopathy, providing valuable insights for applications in preventative eye care and treatment optimization. Through this research, we aim to enhance the utility of AI in clinical settings, potentially transforming early diagnostic processes for ocular diseases. This work introduces a lightweight CNN architecture constructed using a bottom-up approach specifically for automated exudate detection in retinal images. To bolster the model's adaptability to real-world clinical environments, novel exudate-specific data augmentation techniques are proposed. These augmentations account for both intrinsic and extrinsic environmental factors encountered in optician clinics, aiming to enhance the model’s generalizability to these settings. Building upon prior research \cite{aydin2023domain}, this new architecture incorporates advanced internal filter suppression techniques, reducing its computational load by 1.69 million parameters while simultaneously boosting the F1 score. The enhanced architecture, requiring only 4.73 million learnable parameters as opposed to ResNet-18's 11.69 million, not only meets our lightweight design goals but also demonstrates exceptional diagnostic performance with an overall F1 score of 90\%. This underscores the effectiveness of the proposed data augmentation and regularization strategies, confirming the architecture’s operational efficiency in a clinical context.

The remaining work is organized as follows: Section 2, the Literature Review, delves into prior research relevant to this study, providing a comprehensive examination of developments in the detection of retinal exudates and advancements in convolutional neural network architectures. Section 3, the Methodology, details the dataset used, describes the data augmentation techniques designed to simulate real-world clinical scenarios, and outlines the specifics of the proposed lightweight CNN architecture. Section 4, the Model Evaluation, conducts a thorough analysis of the model's performance, discussing the selection of hyperparameters, the effectiveness of the augmented dataset, and the roles of batch normalization and dropout in enhancing diagnostic accuracy. Section 5, the Discussion, interprets the results, compares them with existing methodologies, and assesses the practical implications of the findings. Section 6, the Conclusion, summarizes the study's key contributions, discusses potential limitations, and suggests directions for future research in enhancing AI-driven diagnostics in ophthalmology.

\section{Literature Review}
The application of image segmentation in the detection of retinal diseases represents a pivotal area of research within medical imaging and diagnostic technology. This review explores a range of innovative methodologies that have significantly advanced the ability to detect and analyze pathological features in retinal fundus images. By integrating advanced machine learning techniques with traditional imaging processes, recent developments have markedly enhanced both the accuracy and efficiency of diagnosing retinal conditions. These technological evolutions, particularly in how they address challenges related to feature extraction, classification, and the generalizability of findings across diverse patient populations, underscore the dynamic nature of this research area.

Building on the potential of image segmentation for retinal disease detection, Tang et al.\cite{tang2012splat} introduced the 'splat-feature' classification mechanism. This supervised learning approach partitions retinal fundus images into non-overlapping regions called 'splats', consisting of groups of pixels with similar color and spatial characteristics, potentially crucial for capturing local image features relevant to disease identification. After feature extraction from these splats, a wrapper approach is employed for optimal feature selection, and the selected features are used to train a classifier. Tang et al. achieved a remarkable area-under-the-curve (AUC) of 96\%, demonstrating the method’s promise for automated retinal disease detection. Similarly, Tan et al. \cite{tan2017automated} proposed a two-stage deep learning detector to categorize microaneurysms (MAs), hemorrhages, and exudates in retinal images. Their approach, which utilizes class segmentation for classification, achieved a sensitivity of 71.58\% for the exudate category. However, they acknowledge limitations in exudate detection due to significant variability in their shape and size, posing challenges for pixel-wise segmentation techniques.Further extending the use of segmentation classifiers, Guo et al. \cite{guo2022lighteyes} focused on developing a lightweight CNN architecture designed to optimize computational efficiency while robustly segmenting MAs, exudates, and blood vessels across five publicly accessible datasets. Their innovative architecture facilitates the learning of high-resolution image representations, thereby minimizing the impact on inference speed and maintaining computational efficiency.

Building upon the potential of Deep Learning (DL) for retinal image analysis, Shan et al.\cite{shan2016deep} proposed an innovative methodology utilizing stacked sparse autoencoders (SSAEs) to identify microaneurysms (MAs) in the retina. Their approach involves the SSAE architecture, which facilitates the extraction of latent features from patches of retinal images. These features are then used to distinguish between true and false MAs, achieving a noteworthy AUC of 96.2\%. This performance underscores the potential of SSAEs for robust MA detection in retinal images. In a similar vein, Romero et al. \cite{rosas2015method} introduced a unique method for detecting MAs, employing a 'Bottom-Hat' transformation to enhance the visibility of the reddish regions of MAs while suppressing blood vessels. This technique is further refined through the use of Radon transforms and principal component analysis (PCA) for more robust MA identification. They reported an impressive classification accuracy of 95.93\%, highlighting the clinical applicability of their method. Conversely, Habib et al. \cite{habib2017detection} explored the efficacy of an automated system for MA detection and classification using Gaussian-matched filters for feature extraction, followed by an ensemble classifier for the final decision-making process. However, their approach encountered challenges, as evidenced by a lower-than-expected receiver operating characteristic (ROC) curve with an AUC of only 41.5\%, indicating limitations in the model's ability to generalize during training. Expanding the scope of retinal disease analysis, Zhou et al. \cite{zhou2020benchmark} developed a novel collaborative learning architecture that incorporates attention mechanisms to enhance the segmentation and grading of diabetic retinopathy lesions. This method dynamically enriches annotations at the image level with class-specific information, achieving a promising average AUC for precision and recall of 70.44\%. Their findings underscore the potential of collaborative architectures in advancing the analysis of diabetic retinopathy.

Building further on this theme, Huang et al. \cite{huang2022rtnet} identified a critical gap in the field, noting that prior research primarily focused on architectural design while often overlooking the underlying disease mechanisms. To address this, they proposed a novel approach utilizing a relational transformer block, which, similar to the work of Zhou et al., employs attention mechanisms to effectively characterize the global interdependencies among lesion features and their interactions with vessel features. Despite the exceptional accuracy provided by vision transformers, the substantial volume of training data and significant computational resources required by Huang et al.'s transformer-based architecture could limit its practical application on CPU-based systems, unless effective pruning techniques are employed. Concurrently, Abdullah et al. \cite{abdullah2016localization} addressed the crucial task of detecting and segmenting the optic disc in retinal images. Their methodology combines morphological operations, the grow-cut algorithm, and the Hough transform to enhance the visibility of the optic disc while suppressing background features, particularly the retinal vasculature. Subsequent use of the Hough transform approximates the core of the optic disc, and the grow-cut methodology is applied for precise boundary segmentation. The effectiveness of their approach was validated across five publicly accessible datasets, where it demonstrated exceptional performance, achieving 100\% accuracy on three datasets and exceeding 99\% accuracy on the remaining two. Furthermore, Murugan et al. \cite{murugan2022micronet} introduced a novel three-stage method for autonomous microaneurysm detection in retinal images, encompassing data preprocessing, candidate region identification, and pixel-wise classification. Integrating a CNN architecture with a majority voting classifier, their method demonstrated promising performance, achieving an AUC of 92\%, which is comparable to the state-of-the-art models in this domain.

Gulshan et al.\cite{gulshan2016development} investigated the potential of deep convolutional neural networks (DCNNs) for the automated detection of diabetic retinopathy (DR) in retinal fundus images. Their research focused on using deep learning algorithms to automate the identification of both DR and diabetic macular edema. The DCNN's performance, particularly its sensitivity and specificity for detecting moderate or worse DR, was benchmarked against a reference standard set by a panel of ophthalmologists. Employing a large dataset annotated with varying grades of disease severity, their approach yielded notable results, achieving a sensitivity of 96.5\% and a specificity of 92.4\% in DR detection. Similarly, Rakhlin et al.\cite{rakhlin2017diabetic} utilized DCNNs to develop a method for DR detection, training CNN models on a publicly available Kaggle dataset and evaluating them against the Messidor-2 reference standard. Their CNN model demonstrated high accuracy, with a sensitivity of 99\%, specificity of 71\%, and an AUC of 0.97 when evaluated on the Messidor-2 dataset. These results not only showcased the model's high precision but also highlighted its performance as exceeding that of trained optometrists in DR screening programs.

Reflecting on these studies, the exploration of deep learning applications in retinal disease detection has predominantly focused on employing segmentation techniques to isolate and subsequently classify regions of interest. While this method offers significant advantages in identifying defects, the computational demands of these segmentation architectures and the limitations posed by the availability of large and diverse retinal image datasets warrant further examination. Although datasets like DIARETDB1\cite{kauppi2007diaretdb1} provide accessible resources, their restricted size and class distribution can pose challenges. For example, DIARETDB1 comprises only 89 images, with a scant representation of exudates in just 41 images. Recent research has started to address these challenges by generating synthetic datasets\cite{de2020deep, diaz2019retinal}. Nevertheless, a considerable portion of ongoing scholarly work continues to rely on computationally intensive segmentation and feature extraction techniques for dataset creation\cite{costa2017end}, indicating a clear need for innovation in data handling and algorithm efficiency within the field.

\section{Methodology}
The study employed the publicly accessible EYEPACS dataset from Kaggle \cite{kaggle2015diabetic}, which is widely used for benchmarking diabetic retinopathy (DR) detection algorithms. The dataset comprises retinal fundus images, each annotated with a DR severity rating on a scale from 0 to 4, where '0' indicates no DR, '1' mild DR, '2' moderate DR, '3' severe DR, and '4' proliferative DR. Given the typical challenges associated with class imbalance and repetitive image content in medical datasets, the EYEPACS data underwent a strategic restructuring. Specifically, the images were reclassified into two broad categories to facilitate binary classification: normal (representing DR stage 0) and exudate (encompassing DR stages 1 to 4). To enhance the representativeness and diversity of the training data, a curated subset of 500 unique images was meticulously selected from the original pool. This selection process was guided by the need to ensure an equitable distribution across the new categories, thereby mitigating issues related to class imbalance and sampling redundancy.

\subsection{Dataset}
This study utilized a dataset consisting of 500 retinal fundus images, categorized into two distinct classes: Normal DR images and Exudate DR images. Each participant contributed two images, capturing their left and right eyes, respectively, with both images maintaining identical resolutions. The dataset's composition is detailed in Table  \ref{tab:original-dataset}.

\begin{figure}[!t]
    \centerline{\includegraphics[width=0.7\textwidth]{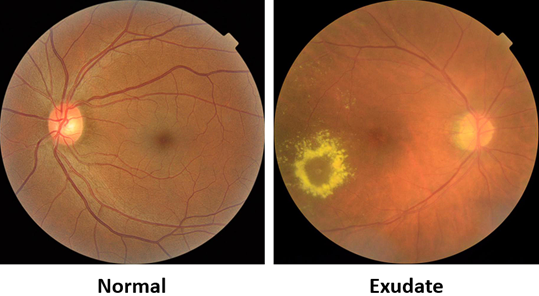}}
    \caption{Normal and Exudate Class Illustrations}
    \label{figure1}
\end{figure}

\begin{table}[htbp]
    \centering
    \caption{Primary Dataset}
    \label{tab:original-dataset}
    \setlength{\tabcolsep}{10pt}
    \renewcommand{\arraystretch}{1.3}
    \begin{tabular}{cc}
        \toprule
        \textbf{Class} & \textbf{Images} \\
        \midrule
        Normal & 250 \\
        Exudate & 250 \\
        \bottomrule
    \end{tabular}
\end{table}
To illustrate the visual differences between the Normal and Exudate classes, Figure \ref{figure1} presents representative images from the dataset. Exudates are characterized by their distinct yellowish coloration and punctate appearance, which differentiate them from other retinal features such as the optic disc, macula, and retinal vasculature. This visual distinction is crucial for guiding the development of data augmentation strategies.

Analyzing fundus images poses significant challenges due to considerable inter-subject variability, which stems from a combination of intrinsic and extrinsic factors. Intrinsic factors include the stage of DR, variations in retinal pigmentation, and intra-retinal contrast differences. Extrinsic factors encompass limitations of the image acquisition hardware, angles of image capture, and patient positioning, which can all contribute to variations in image quality, such as brightness inconsistencies, hardware-induced blur, and textural heterogeneity. Addressing these challenges requires robust image processing techniques to normalize and enhance the dataset for more accurate analysis.

The dataset was divided into training, validation, and testing subsets to support the model development process. This segmentation facilitated thorough training and validation of the proposed architecture’s performance. The partitioning of the dataset is as follows: 70\% for training, 20\% for validation, and 10\% for testing, as detailed in Table \ref{tab:dataset-split}.

\begin{table}[htbp]
    \centering
    \caption{Primary Dataset Distribution}
    \label{tab:dataset-split}
    \begin{tabular}{cccc} 
        \toprule
        & \textbf{Total images} & \textbf{Normal} & \textbf{Exudate} \\
        \midrule
        \textbf{Training Data} & 350 & 175 & 175 \\
        \textbf{Validation Data} & 100 & 50 & 50 \\
        \textbf{Testing Data} & 50 & 25 & 25 \\
        \midrule
        \textbf{Total} & 500 & 250 & 250 \\
        \bottomrule
    \end{tabular}
\end{table}

\subsection{Data Augmentations}
In this study, a series of data augmentation techniques were applied to the original dataset to enhance its diversity and robustness. These augmentations were carefully selected to prepare the proposed architecture for effectively handling various experimental scenarios encountered throughout the research. Each augmentation technique was chosen based on specific considerations aimed at improving the model's performance by simulating a wide range of real-world conditions. The rationale behind these selections is detailed in the following sections.

\subsubsection{Oriented Based Scaling}
\begin{figure}[H]
\centering
\includegraphics[width=0.6\textwidth]{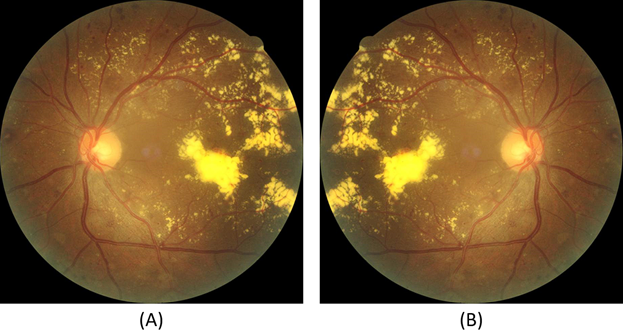}
\label{figure2}
\caption{Horizontal Augmentations: (A) Pre-augmentation, (B) Post-augmentation}
\end{figure}

To bolster the architectural resilience against variations in image orientation, horizontal augmentation was systematically applied to the original dataset. This technique introduces controlled alterations in image orientation, addressing potential discrepancies that may arise during image acquisition, such as variations in eye positioning or deviations caused by imaging equipment. Vertical augmentation was deliberately omitted to avoid the creation of redundant images, which could skew the model's learning process. As depicted in Figure \ref{figure2}, this approach was employed given the common occurrence of fundus images being captured at different horizontal angles by various practitioners during clinical screenings. Horizontal augmentation effectively normalizes these orientation disparities, enhancing the model’s capability to generalize across images with varying angles of capture, thus contributing significantly to the model’s adaptability and robustness.

\begin{figure}[H]
\centering
\includegraphics[width=0.7\textwidth]{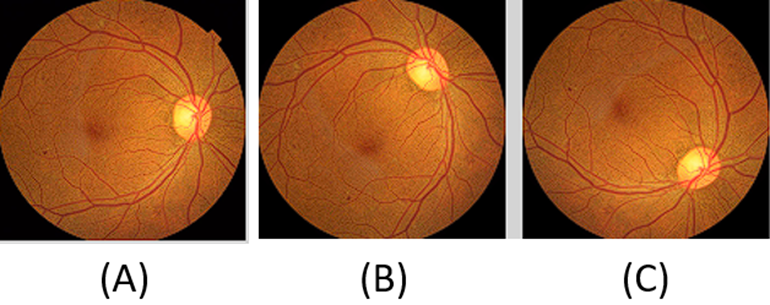}
\caption{Rotation Augmentation at 45\textdegree: (A) Pre-augmentation, (B, C) Post-augmentation}
\label{figure3}
\end{figure}

\begin{figure}[H]
\centering
\includegraphics[width=0.7\textwidth]{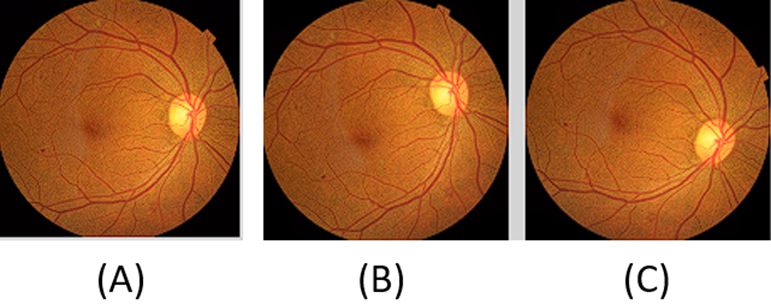}
\caption{Rotation Augmentation at 20\textdegree: (A) Pre-augmentation, (B, C) Post-augmentation}
\label{figure4}
\end{figure}

Figure \ref{figure3} and Figure \ref{figure4} demonstrate the application of rotation-based augmentations, featuring rotations of 45\textdegree and 20\textdegree, respectively. The motivation for incorporating rotations into the dataset is to mimic the varied angles at which retinal examinations are conducted, reflecting the diversity of viewpoints an optometrist might adopt in clinical practice. This augmentation strategy is designed to enhance the model’s ability to recognize and accurately classify retinal lesions from different orientations, thereby reducing the risk of overfitting. Since the dataset includes images captured by various optometrists, it naturally contains examples taken at diverse angles. These rotational augmentations further enrich the dataset, ensuring that the model develops a robust ability to generalize across a wide range of real-world imaging conditions.

Here's an updated and more technically detailed version of your discussion on contrast variability and the subsequent augmentation processes:

\subsubsection{Contrast Variability}
The acquisition of diabetic retinopathy (DR) images often occurs under a variety of environmental conditions, which can introduce significant variability in room illumination, camera specifications, and image capture angles. To equip the proposed architecture with the capability to adapt to such variability, the training dataset underwent contrast modifications, which included adjustments to brightness and blur levels.

\begin{figure}[H]
\centering
\includegraphics[width=0.7\textwidth]{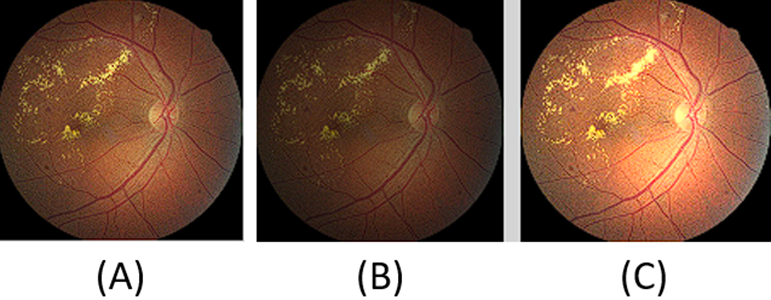}
\caption{Pixel Brightness Augmentation at 38\textdegree: (A) Pre-augmentation, (B, C) Post-augmentation}
\label{figure5}
\end{figure}

\begin{figure}[H]
\centering
\includegraphics[width=0.7\textwidth]{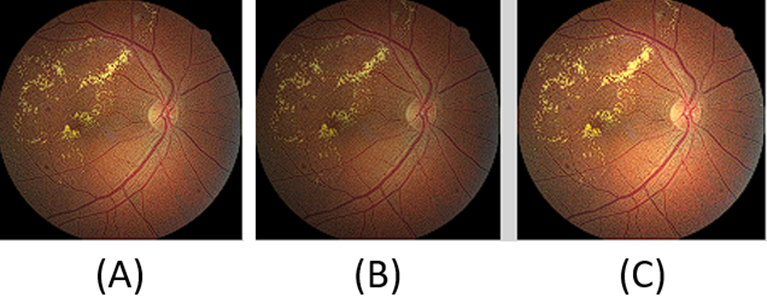}
\caption{Pixel Brightness Augmentation at 20\textdegree: (A) Pre-augmentation, (B, C) Post-augmentation}
\label{figure6}
\end{figure}

Figures \ref{figure5} and \ref{figure6} illustrate the application of brightness augmentation at angles of 38\textdegree and 20\textdegree, respectively. This augmentation technique was specifically designed to simulate and counteract the fluctuating lighting conditions that frequently occur during the image acquisition process in diverse clinical environments. By enhancing the model's adaptability to different lighting conditions, brightness augmentation significantly improves its generalizability across various clinical settings.

\begin{figure}[H]
\centering
\includegraphics[width=0.6\textwidth]{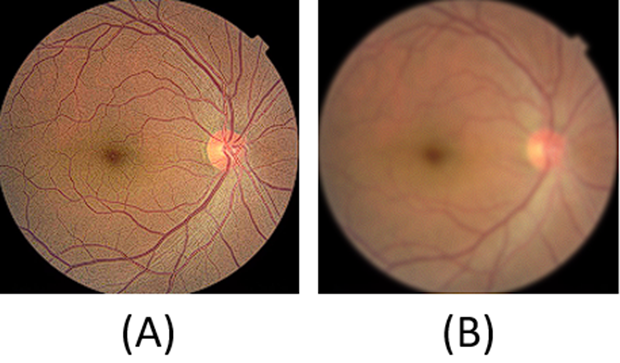}
\caption{Blur Augmentation: (A) Pre-augmentation, (B) Post-augmentation}
\label{figure7}
\end{figure}

Figure \ref{figure7} showcases the implementation of blur augmentation using a 3px Gaussian blur. This augmentation aims to mimic the potential blurring effects caused by camera lens imperfections or suboptimal focusing during the capture process. Incorporating blur augmentation allows the architecture to effectively process and recognize features in images that exhibit similar blurring characteristics, thus enhancing its diagnostic accuracy in real-world scenarios.

After applying the aforementioned data augmentation techniques, the dataset expanded to a total of 3286 images. Table \ref{tab:augmented-dataset} provides a detailed breakdown of the augmented dataset, indicating an equitable distribution across the two classes.

\begin{table}[htbp]
    \centering
    \caption{Distribution of the Augmented Dataset}
    \label{tab:augmented-dataset}
    \setlength{\tabcolsep}{10pt}
    \renewcommand{\arraystretch}{1.3} 
    \begin{tabular}{cc} 
        \toprule
        \textbf{Class} & \textbf{Images} \\
        \midrule
        Normal & 1,615 \\
        Exudate & 1,671 \\
        \bottomrule
    \end{tabular}
\end{table}

The augmented dataset was then stratified into training, validation, and testing subsets, maintaining the original proportions of data distribution to ensure a balanced representation across all classes. This strategic partitioning facilitates a robust and comprehensive evaluation of the model. Detailed proportions of the dataset division are presented in Table  \ref{tab:augmented-dataset-split}. with the training, validation, and testing sets comprising 70\%, 20\%, and 10\% of the total dataset, respectively.

\begin{table}[htbp]
    \centering
    \caption{Augmented Dataset Split}
    \label{tab:augmented-dataset-split}
    \begin{tabular}{lccc}
        \toprule
        & \textbf{Total} & \textbf{Normal} & \textbf{Exudate} \\
        \midrule
        \textbf{Training Data} & 2,300 & 1,104 & 1,196 \\
        \textbf{Validation Data} & 658 & 325 & 333 \\
        \textbf{Testing Data} & 328 & 186 & 142 \\
        \midrule
        \textbf{Total} & 3,286 & 1,615 & 1,671 \\
        \bottomrule
    \end{tabular}
\end{table}

\subsection{Proposed Architecture}
To achieve a lightweight architecture suitable for deployment on constrained hardware environments, the design strategy focused on incorporating dual convolutional blocks. Each block utilizes a limited number of filters followed by a max-pooling layer, facilitating efficient feature extraction. These feature maps are then processed through a ReLU activation function to introduce necessary non-linearity into the network. The selection of ReLU, defined by Equation \ref{eq:1}, was driven by its computational efficiency relative to sigmoid and tanh functions.

\begin{equation}
    A(z) = \max(0, z) 
    \label{eq:1}
\end{equation}

The neural network culminates in a two-layer fully connected (FC) network, with the first FC layer comprising 100 neurons and the second containing 40 neurons. Notably, the convolutional blocks utilize a sparse arrangement of filters (9 and 18 respectively) to minimize computational demand. While this streamlined approach reduces complexity, it necessitates robust feature extraction capabilities, hence the strategic integration of tailored data augmentations to enhance generalization potential.

\begin{figure}[H]
    \centering
    \includegraphics[width=0.8\textwidth]{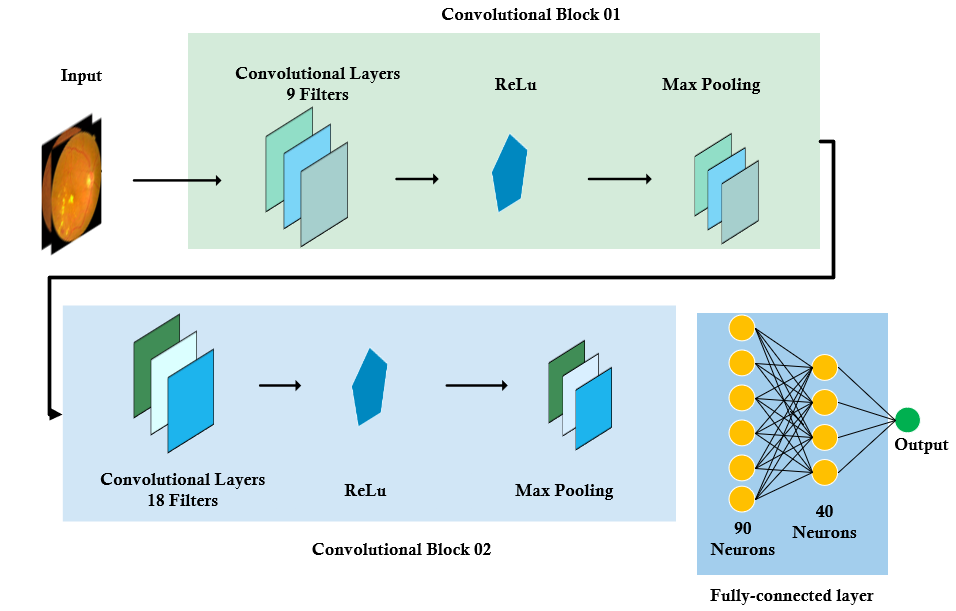}
    \caption{Proposed Architecture}
    \label{figure8}
\end{figure}

Additionally, the architecture incorporates regularization strategies like batch normalization and dropout within its internal layers to enhance training dynamics and prevent overfitting. Batch normalization, as outlined by Equations \ref{eq:2} to \ref{eq:5}, normalizes layer inputs to stabilize learning and accelerate convergence.

\begin{equation}
    {u}^{[l]} = \frac{1}{m} \sum_{i} k^{[l](i)}
    \label{eq:2}
\end{equation}

\begin{equation}
    \sigma^{[l]2} = \frac{1}{m} \sum_{i} (k^{[l](i)} - u^{[l]})^2
    \label{eq:3}
\end{equation}

\begin{equation}
    k_{\text{norm}}^{[l](i)} = \frac{k^{[l](i)} - u^{[l]}}{\sqrt{\sigma^{[l]2} + \epsilon}}
    \label{eq:4}
\end{equation}

\begin{equation}
    \underline{k}_{\text{norm}}^{[l](i)} = \gamma^{[l]} \underline{k}_{\text{norm}}^{[l](i)} + \beta^{[l]}
    \label{eq:5}
\end{equation}

The architecture's layout is meticulously detailed in Table \ref{tab:architecture-layout}, illustrating its efficient configuration that only comprises 4.73 million parameters—substantially fewer than more complex models like ResNet-18, which possesses 11.69 million parameters.

\begin{table}[htbp]
    \centering
    \caption{Internal Block-wise Architecture of the Model}
    \label{tab:architecture-layout}
    \begin{tabular}{lcc}
        \toprule
        \textbf{Layers} & \textbf{Output Shape} & \textbf{Parameters} \\
        \midrule
        Input & 3, 224x224 & --- \\
        Conv-1 & 9, 222x222 & 1.73K \\
        BatchNorm2d & 9, 222x222 & 36 \\
        ReLU & 9, 222x222 & --- \\
        Max-Pool & 9, 111x111 & --- \\
        Conv-2 & 18, 109x109 & 1.48K \\
        BatchNorm2d & 18, 109x109 & 72 \\
        ReLU & 18, 109x109 & --- \\
        Max-Pool & 18, 54x54 & --- \\
        FullyConv1 & 90 Neurons & 4,720K \\
        ReLU & 90 & --- \\
        FullyConv2 & 40 Neurons & 3.64K \\
        ReLU & 40 & --- \\
        Output & 2 Neurons & 82 \\
        \midrule
        \textbf{Total Parameters} & --- & 4.73 Million \\
        \bottomrule
    \end{tabular}
\end{table}

\section{Model Evaluation}
\subsection{Hyperparameter Tuning}
To optimize the architecture for constrained computational resources, the model's design, development, training, and validation were conducted using Google Colab, leveraging the computational power of the Google Cloud Platform. The development framework utilized was PyTorch. For ensuring consistent and unbiased comparisons across different experiments, a set of global hyperparameters were predefined and uniformly applied across all training sessions. These hyperparameters are summarized in Table \ref{tab:global-hyperparameters} for easy reference.

\begin{table}[htbp]
    \centering
    \caption{Global Hyperparameters}
    \label{tab:global-hyperparameters}
    \begin{tabular}{cc}
        \toprule
        \textbf{Hyperparameter} & \textbf{Value} \\
        \midrule
        Batch Size & 32 \\
        Epochs & 40 \\
        Learning Rate & 0.02 \\
        Optimizer & SGD with momentum (SGD-M) \\
        Loss Function & Cross-Entropy Loss \\
        \bottomrule
    \end{tabular}
\end{table}

\subsection{Original Dataset Performance}
The initial performance assessment of the model utilized the original, unaugmented dataset. The performance metrics derived from this experiment are detailed in Table \ref{tab:performance-original-dataset}, illustrating the limitations in generalizability with an F1-score of 79\%. This underperformance was partially attributed to the relatively small dataset size, which contained only 500 images. To mitigate this and enhance the model's generalizability, data augmentation was strategically employed rather than increasing the model's complexity, which would escalate the computational overhead significantly.

\begin{table}[htbp]
    \centering
    \caption{Performance of Original Dataset}
    \label{tab:performance-original-dataset}
    \begin{tabular}{cc}
        \toprule
        \textbf{Metric} & \textbf{Value} \\
        \midrule
        Precision & 69\% \\
        Recall & 92\% \\
        F1-Score & 79\% \\
        Accuracy & 83\% \\
        \bottomrule
    \end{tabular}
\end{table}

\begin{figure}[H]
    \centering
    \includegraphics[width=0.5\textwidth]{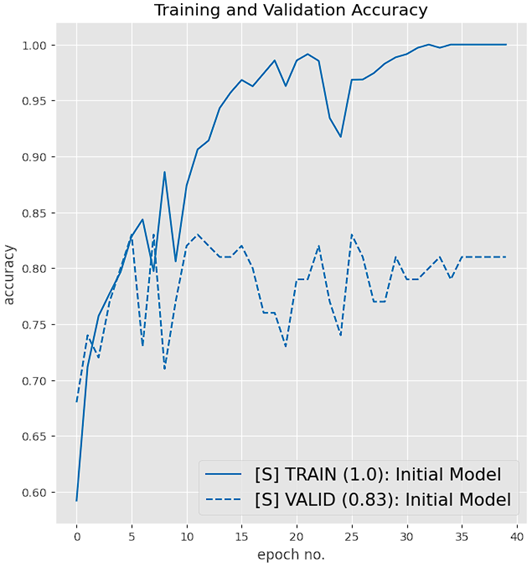}
    \caption{Validation Accuracy over Epochs for Original Dataset}
    \label{figure9}
\end{figure}

Despite these adjustments, the validation accuracy plateaued at a modest 83\%, as depicted in Figure \ref{figure9}. This plateau further underscores the challenges posed by the limited dataset size and variability.

\begin{figure}[H]
    \centering
    \includegraphics[width=0.4\textwidth]{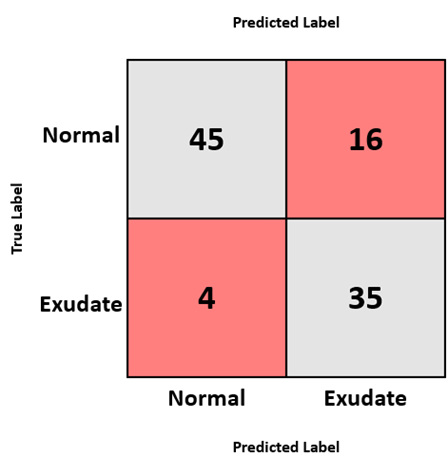}
    \caption{Confusion Matrix of Original Dataset}
    \label{figure10}
\end{figure}

To further illuminate the training and validation dynamics, Figure \ref{figure10} provides the confusion matrix for the original dataset. The matrix visually represents the class-specific breakdown of classification accuracy, highlighting instances where 16 normal fundus samples were incorrectly classified as containing exudates, and 4 exudate samples were misclassified as normal.
\subsection{Augmented Dataset Performance}
To minimize computational complexity while enhancing model performance, data augmentation was favored over increasing the convolutional depth of the architecture. This strategic choice is validated by the observed enhancements in model metrics, as shown in Table \ref{tab:performance-augmented-dataset}, where precision and F1-score have significantly improved.

\begin{table}[htbp]
    \centering
    \caption{Augmented Dataset Performance}
    \label{tab:performance-augmented-dataset}
    \begin{tabular}{cc}
        \toprule
        \textbf{Metric} & \textbf{Value} \\
        \midrule
        Precision & 81\% \\
        Recall & 90\% \\
        F1-Score & 85\% \\
        Accuracy & 85\% \\
        \bottomrule
    \end{tabular}
\end{table}

\begin{figure}[H]
    \centering
    \includegraphics[width=0.5\textwidth]{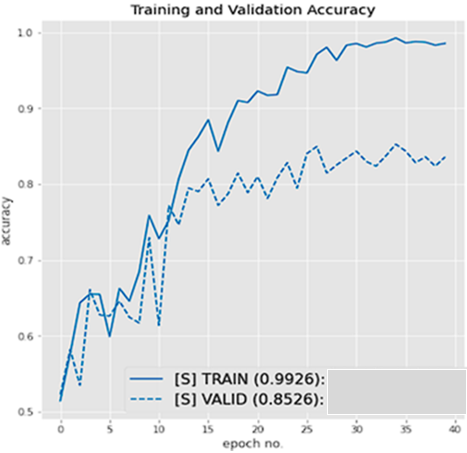}
    \caption{Performance of Augmented Dataset}
    \label{figure11}
\end{figure}

As illustrated in Figure \ref{figure11}, training the architecture with the augmented dataset led to a moderate improvement, achieving an overall F1-score of 85\%. This increment underscores the potential of data augmentation to boost model capabilities, although alone it may not suffice for achieving optimal performance. Subsequently, advanced regularization techniques such as batch normalization and dropout were integrated to further refine the architecture’s performance.

\begin{figure}[H]
    \centering
    \includegraphics[width=0.5\textwidth]{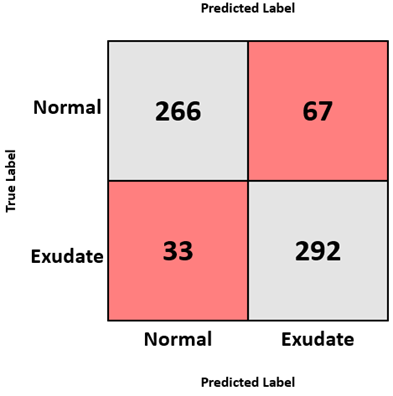}
    \caption{Confusion Matrix of Augmented Dataset}
    \label{figure12}
\end{figure}

Moreover, the confusion matrix in Figure \ref{figure12} elucidates misclassification trends within the augmented dataset. Notably, 67 samples originally classified as normal were misidentified as exudates, while 33 exudate samples were incorrectly labeled as normal, illustrating areas where further model tuning might be beneficial.

\subsection{Batch Normalization}
Despite the initial improvements achieved with data augmentation, the count of misclassifications remained relatively high. Consequently, subsequent experiments focused on incorporating batch normalization to optimize the architecture's generalization capacity by addressing internal covariance among classes.

\begin{figure}[H]
    \centering
    \includegraphics[width=0.5\textwidth]{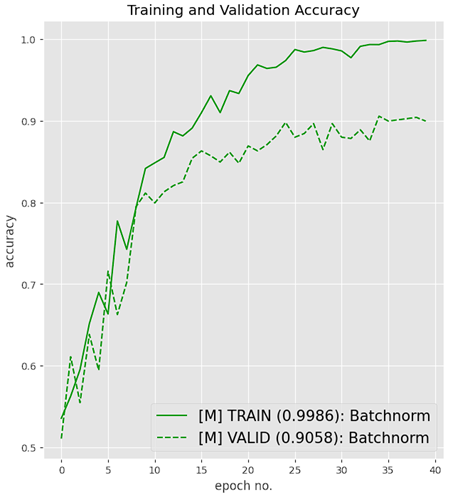}
    \caption{Performance Enhancement with Batch Normalization}
    \label{figure13}
\end{figure}

As depicted in Table \ref{tab:performance-batchnorm}, the inclusion of batch normalization significantly improved all key performance metrics. Precision increased to 88\%, recall to 93\%, and the F1-score peaked at 90.6\%. Figure \ref{figure13} visualizes these enhancements, showing the augmented dataset's performance when trained on the proposed architecture with batch normalization.

\begin{table}[htbp]
    \centering
    \caption{Performance of Proposed Architecture with Batch Normalization}
    \label{tab:performance-batchnorm}
    \begin{tabular}{cc}
        \toprule
        \textbf{Metric} & \textbf{Value} \\
        \midrule
        Precision & 88\% \\
        Recall & 93\% \\
        F1-Score & 90.6\% \\
        Accuracy & 90.6\% \\
        \bottomrule
    \end{tabular}
\end{table}

\begin{figure}[H]
    \centering
    \includegraphics[width=0.5\textwidth]{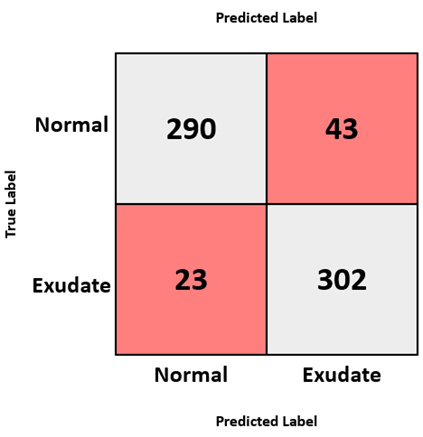}
    \caption{Confusion Matrix Post-Batch Normalization}
    \label{figure14}
\end{figure}

Further analysis of the confusion matrices before and after the application of batch normalization (Figures \ref{figure12} and \ref{figure14}) shows a significant reduction in misclassifications. The number of normal fundus images erroneously classified as exudate decreased from 67 to 43, underscoring the efficacy of batch normalization in enhancing diagnostic precision.

\subsection{Dropout}
To mitigate overfitting, a subsequent experiment incorporated dropout, a technique that randomly deactivates neurons during training to prevent the model from overly depending on specific neuron connections. This promotes robustness and generalizability across unseen data. The optimal dropout rate was determined through an iterative process, which involved incremental adjustments and evaluating the impact on metrics indicative of overfitting.

Table \ref{tab:dropout-comparison} showcases the effects of varying dropout rates on model performance, illustrating changes in F1 scores, training and validation accuracies, and the degree of overfitting. Notably, increasing the dropout rate generally decreases overfitting, with minor variations in performance metrics such as precision and recall.

\begin{table*}[htbp]
    \centering
    \caption{Performance Comparison of the Proposed Architecture with Varied Dropout Rates}
    \label{tab:dropout-comparison}
    \begin{tabular}{cccccc}
        \toprule
        \textbf{Dropout Rate} & \textbf{Training Accuracy} & \textbf{Validation Accuracy} & \textbf{Degree of Overfitting} & \textbf{F1-Score} \\
        \midrule
        30\% & 98.39\% & 88.15\% & 10.24\% & 88\% \\
        40\% & 97.90\% & 88.45\% & 9.45\% & 88\% \\
        50\% & 97.35\% & 89.36\% & 7.99\% & 89\% \\
        60\% & 83.66\% & 77.66\% & 6.0\% & 73\% \\
        70\% & 73.77\% & 75.53\% & -1.76\% & 74\% \\
        \bottomrule
    \end{tabular}
\end{table*}

Further analysis, summarized in Table \ref{tab:performance-dropout}, identified a dropout rate of 0.5 as optimal, balancing precision and recall while achieving an F1-score of 89\%. This setting resulted in a consistent performance across various metrics.

\begin{table}[htbp]
    \centering
    \caption{Performance Evaluation of the Proposed Architecture with 0.5 Dropout Rate}
    \label{tab:performance-dropout}
    \begin{tabular}{cc}
        \toprule
        \textbf{Metric} & \textbf{Value} \\
        \midrule
        Precision & 88\% \\
        Recall & 90\% \\
        F1-Score & 89\% \\
        Accuracy & 89.36\% \\
        \bottomrule
    \end{tabular}
\end{table}

\begin{figure}[H]
    \centering
    \includegraphics[width=0.5\textwidth]{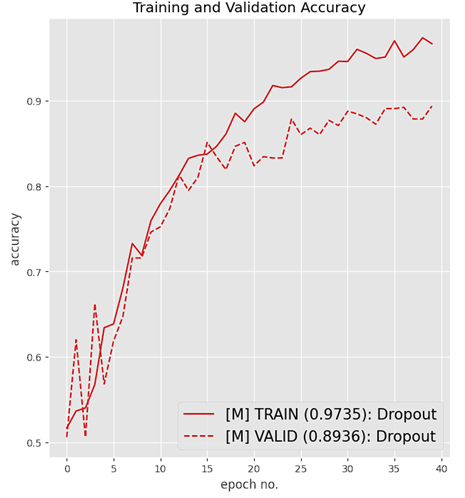}
    \caption{Modified Architecture Performance with Dropout}
    \label{figure15}
\end{figure}

\begin{figure}[H]
    \centering
    \includegraphics[width=0.5\textwidth]{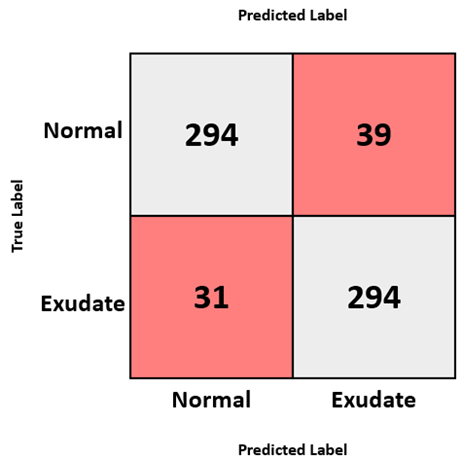}
    \caption{Dropout Confusion Matrix}
    \label{figure16}
\end{figure}

Figures \ref{figure15} and \ref{figure16} illustrate the performance and confusion matrix for the model with a 0.5 dropout rate, respectively, demonstrating a marked improvement in the classification of normal fundus images compared to previous setups.

\subsection{Batch Normalization and Dropout}
Encouraged by the individual improvements observed with dropout in precision and F1-score, and batch normalization in recall, the final experiment sought to investigate the synergistic effects of combining these techniques within the proposed architecture. This experiment aimed to determine if their combined application could yield further performance enhancements.
\begin{table}[htbp]
    \centering
    \caption{Evaluation of the Proposed Architecture Utilizing 0.5 Dropout and Batch Normalization}
    \label{tab:performance-results}
    \begin{tabular}{cc}
        \toprule
        \textbf{Metric} & \textbf{Value} \\
        \midrule
        Accuracy & 68.69\% \\
        Precision & 76\% \\
        Recall & 53\% \\
        F1-Score & 63\% \\
        \bottomrule
    \end{tabular}
\end{table}

However, the integration of both batch normalization and a 0.5 dropout ratio resulted in a detrimental effect, particularly on recall, which dropped to its lowest value of 53\%. This decline might be attributed to the inherent simplicity of the architecture, which includes only two convolutional blocks. The strong combined regularization effects from both batch normalization and dropout might have overly constrained the model's ability to generalize effectively to the dataset.

\begin{figure}[H]
    \centering
    \includegraphics[width=0.5\textwidth]{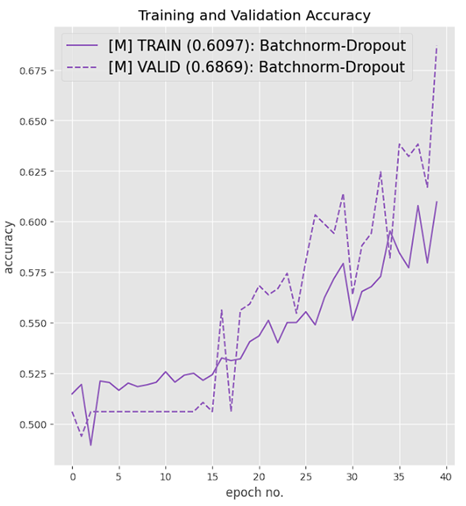}
    \caption{Proposed Architecture with Batch-norm and Dropout}
    \label{figure17}
\end{figure}

A comparative analysis of experimental performances across different setups is detailed in Table \ref{tab:experiment-performance}, highlighting the various impacts of each adjustment made during the experimentation phase.

\begin{table*}[htbp]
    \centering
    \caption{Comparison of Experimental Performance}
    \label{tab:experiment-performance}
    \begin{tabular}{lcccc}
        \toprule
        \textbf{Experimental Setup} & \textbf{Precision} & \textbf{Recall} & \textbf{F1 Score} & \textbf{Validation Accuracy} \\
        \midrule
        Original Dataset & 69\% & 92\% & 79\% & 83\% \\
        Augmented Dataset & 81\% & 90\% & 85\% & 85.26\% \\
        Batch-norm & 88\% & 93\% & 90\% & 90.58\% \\
        Dropout 50\% & 88\% & 90\% & 89\% & 89.36\% \\
        Batch-norm with Dropout 50\% & 76\% & 53\% & 63\% & 68.69\% \\
        \bottomrule
    \end{tabular}
\end{table*}

\begin{figure}[H]
    \centering
    \includegraphics[width=0.5\textwidth]{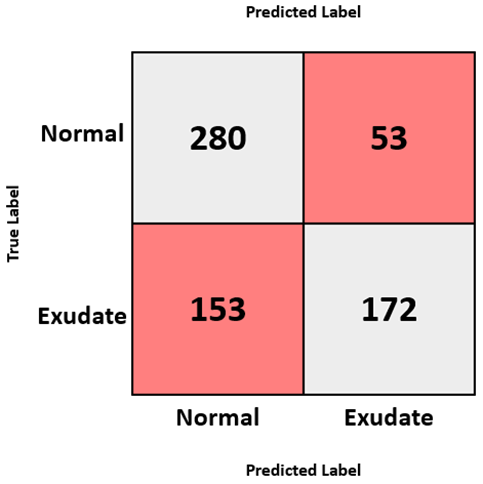}
    \caption{Confusion Matrix for Batch Normalization and Dropout}
    \label{figure18}
\end{figure}

The confusion matrix in Figure \ref{figure18} reveals the specific misclassification trends, where 53 normal fundus images were incorrectly identified as exudates, and 153 exudate samples were mislabeled as normal. This indicates significant false positives and false negatives, illustrating the challenge faced by the architecture in maintaining a precise classification balance between normal and exudate samples.

\section{Discussion}

To systematically determine the optimal architecture for detecting exudates, this study adopted a methodical development pipeline featuring incremental enhancements. Rather than presenting only the final model, a phased training and validation approach was utilized. Each design modification was introduced sequentially, with the system undergoing comprehensive training and validation to evaluate its impact on performance. This iterative methodology started with the original dataset as the baseline.

\begin{figure*}[h]
    \centering
    \includegraphics[width=1\textwidth]{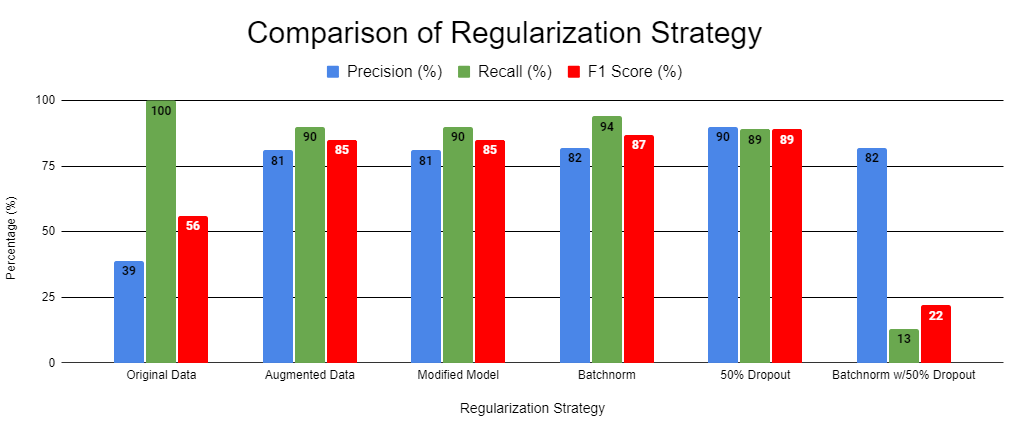}
    \caption{Evaluation of Experimental Strategies}
    \label{figure19}
\end{figure*}

The initial phase involved the application of domain-specific data augmentations, followed by an assessment of various regularization techniques. Batch normalization proved to be the most effective, enhancing the overall F1-score to 90\%. In contrast, implementing dropout at a rate of 0.5 resulted in a nearly equivalent F1-score of 89\%, albeit with a noticeable trade-off, reducing recall to 90\% while maintaining precision at 88\%. This represents a 3\% decline in recall compared to the results achieved with batch normalization alone.

Figure \ref{figure19} visually delineates the performance improvements across different regularization strategies explored in this study, highlighting the significant benefits of the proposed data augmentation techniques. These techniques adeptly captured the domain-specific characteristics of exudate detection. Compared to the baseline dataset, the augmented dataset significantly boosted all evaluation metrics, achieving a notable 12\% improvement in precision and a 6\% increase in F1 score.

The simultaneous implementation of batch normalization and dropout, however, had a detrimental impact on performance, with the F1 score plummeting to 63\%. This decline is likely due to the limited capacity of the proposed architecture's convolutional blocks. The compounded effect of these robust regularization techniques may have obstructed the model's ability to adequately learn generalizable features.

In our pursuit of crafting a streamlined architecture, Table \ref{tab:computational-comparison} presents a detailed computational comparison between our proposed model, established state-of-the-art architectures, and prior work\cite{aydin2023domain}. This comparison focuses on computational load, particularly in terms of the number of learnable parameters.

\begin{table}[htbp]
    \centering
    \caption{Computational Comparison}
    \label{tab:computational-comparison}
    \begin{tabular}{cc}
        \toprule
        \textbf{Model} & \textbf{Parameters (Millions)} \\
        \midrule
        GoogleNet & 13 \\
        ResNet-18 & 11.69 \\
        Light-weight CNN\cite{aydin2023domain} & 6.42 \\
        Proposed & 4.73 \\
        \bottomrule
    \end{tabular}
\end{table}

A discernible pattern emerges from this analysis. Our innovative architecture markedly stands out for its significantly reduced parameter count, comprising just 4.73 million parameters. This contrasts sharply with the more parameter-intensive ResNet-18, which contains 11.69 million parameters, and even lightweight CNN research \cite{aydin2023domain}, which was based on a 6.42 million parameter framework.

\section{Conclusion}
This research marks a significant achievement in the development of a streamlined Convolutional Neural Network (CNN) architecture tailored for exudate detection in retinal fundus images. With a modest parameter count of 4.73 million, the proposed architecture achieves an impressive F1 score of 90\%. This performance highlights the effectiveness of the enhancements introduced, including domain-specific data augmentations that adapt the data to mirror the characteristics of the application domain closely. These modifications have led to substantial improvements in the model’s predictive accuracy across various scenarios.

The proposed architecture not only facilitates efficient computations but also sets a benchmark for deploying advanced CNN models on hardware with limited computational capacity. Looking ahead, future research will focus on enhancing model accuracy through the integration of custom-defined attention mechanisms. These mechanisms are designed to direct the model’s focus more precisely towards localized abnormalities, thereby improving diagnostic precision. Additionally, to increase the transparency and traceability of decision-making processes, efforts will be made to incorporate saliency mapping. This will provide clearer insights into the underlying reasons for the model’s predictions, addressing a critical aspect of model interpretability.

Moreover, the scalability of the proposed model design allows for its application across various fields that require robust CNN implementations but face computational constraints. These fields include security\cite{hussain2019deployment,alsboui2022dynamic}, manufacturing\cite{hussain2023custom}, food monitoring\cite{hussain2022feature}, and renewable energy\cite{zahid2023lightweight,hussain2023review, animashaun2023automated}. Further research will also explore hyperparameter optimization strategies, such as variations in optimizer configurations, to refine the model’s efficiency and effectiveness even further. This ongoing work underscores our commitment to advancing the capabilities of AI-driven diagnostic tools in medical imaging ~\cite{hussain2023exudate} and beyond, aiming to enhance outcomes and operational efficiency across diverse application domains.

\bibliographystyle{unsrt} 
\bibliography{references}

\end{document}